\documentclass[raa]{raa}            

\usepackage{graphicx,times}             
\usepackage{natbib}
\usepackage{amssymb,amsmath}
\bibpunct{(}{)}{;}{a}{}{,}

\usepackage[a4paper=true,dvipdfm=true,pagebackref=true]{hyperref}
\hypersetup{colorlinks = true, linkcolor = green, anchorcolor = red, citecolor = blue, filecolor = red, pagecolor = red, urlcolor = red}

\begin{document}

   \title{AstroSat observation of the Be/X-ray binary Pulsar 3A 0726-260 (4U 0728-25)
}

   \volnopage{Vol.0 (20xx) No.0, 000--000}      
   \setcounter{page}{1}          

   \author{Jayashree Roy
      \inst{1,2}
   \and P. C. Agrawal
      \inst{1}
   \and  B. Singari
      \inst{3}
   \and R. Misra
     \inst{2}
   }

   \institute{UM-DAE Centre for Excellence in Basic Sciences, University of Mumbai, Vidyanagari Campus, Mumbai, 400098, Maharashtra, India.\\
        \and
             Inter-University Center for Astronomy and Astrophysics, Post Bag 4, Pune, 411007, Maharashtra, India.\\
        \and
             School of Physical Sciences, National Institute of Science Education and Research Bhubaneswar, Khurda, 752050, Odisha, India.\\
\vs\no
   {\small Received~~20xx month day; accepted~~20xx~~month day}}

\abstract{ Results on timing and spectral properties of the Be/X-ray binary pulsar 3A 0726-260 (4U 0728-25) are presented. The binary was observed on 2016 May 6-7 with the Large Area X-ray Proportional Counter (LAXPC) and Soft X-ray Telescope (SXT) instruments onboard the AstroSat satellite. During this observation the source was in non-flaring persistent state at a flux level of $\sim$ 8.6 $\pm$ 0.3 $\times$10$^{-11}$ ergs cm$^{-2}$ sec$^{-1}$ in 0.4-20 keV. Strong X-ray pulsations with a period of 103.144$\pm$0.001 seconds are detected in 0.3-7 keV with the SXT and in 3-40 keV with the LAXPC. The pulse profile is energy dependent, and there is an indication that the pulse shape changes from a broad single pulse to a double pulse at higher energy. At energies above 20 keV, we report the first time detection of pulsation period 103.145$\pm$0.001 seconds and the double peaked pulse profile from the source.
The energy spectrum of the source is derived from the combined analysis of the SXT and LAXPC spectral data in 0.4-20 keV. The best spectral fit is obtained by a power law model with a photon index (1.7$\pm$0.03) with high energy spectral cut-off at 12.9 $\pm$ 0.7 keV. A broad Iron line at $\sim$ 6.3 keV is detected in the energy spectrum. We briefly discuss the implications of these results.}

\email{jayashree@iucaa.in}
   \authorrunning{J. Roy et al. }            
   \titlerunning{AstroSat observation of 3A 0726-260 (4U 0728-25) }  

   \maketitle

%
%
\section{Introduction}           
\label{sect:intro}
 The X-ray source 3A 0726-260 (also known as X 0726-260 and 4U 0728-25) is one of the least studied X-ray binary pulsar. This source was first reported in the Fourth Uhuru catalog \citep{1978ApJS...38..357F}. Further, X-ray detections were reported by ARIEL V \citep{1981MNRAS.197..865W} and HEAO-1 \citep{1984ApJS...56..507W}. From the Scanning Modulation Collimator experiment on HEAO-1 \citet{1984ApJ...280..688S} identified the X-ray source with a B0 type star at a distance of 4.6-6 kpc. The optical spectrum of the star showed broad H$\alpha$  line (13 \AA ~FWHM) in emission and H$\beta$ line in absorption, suggesting it to be a Be binary \citep{1984ApJ...280..688S}. \citet{1984A&A...131..385C} found both H$\alpha$ (5.5 \AA) and weak H$\beta$ (0.4 \AA) emission lines in the spectra of the optical counterpart and revised its spectral type as B0Ve and distance as $\sim$ 4.6$\pm$ 1.6 kpc.  
\citet{1996A&A...315..160N} studied the optical spectra and performed photometry in the optical and infrared region of the Be star and classified it as O8-9Ve type star at a distance of $\sim$ 6.1$\pm$0.3 kpc.
Despite being a persistent X-ray emitter at luminosity level of $\sim$ 10$^{35}$ ergs s$^{-1}$ during non-flaring states, 3A 0726-260 is one of the least studied X-ray binaries. The only reported X-ray study of this binary is by \citet{1997ApJ...489L..83C}, who analyzed its 1996 observations by the All Sky Monitor (ASM) and Proportional Counter Array (PCA) instruments onboard the  Rossi X-ray Timing Explorer (RXTE) satellite.  From the PCA light curves, they discovered that this binary has an X-ray pulsar with a spin period of 103.2$\pm$0.1 sec. They also inferred from the ASM light curve the orbital period of the binary  $\sim$ 34.5 days, and it's harmonic at 17.5 days. The orbital period has been refined to be 34.5446$\pm$0.0153 days by \citet{2016ATel.9820....1N} using MAXI/GSC observations. Using the Swift BAT data \citet{2016ATel.9823....1C} have also reported improved value of the orbital period as 34.548$\pm$0.010 days. \citet{1997ApJ...489L..83C} also studied the energy spectrum of this binary in 2-20 keV from the RXTA-PCA data and found that it is well described by an absorbed power law model. They derived a photon index of 1.58$\pm$0.03 and a hydrogen column density of 0.3$\times$10$^{22}$ cm$^{2}$. Estimated unabsorbed 2-20 keV X-ray flux with this model is 6.9$\times$10$^{-11}$ ergs cm$^{-2}$ sec$^{-1}$.
AstroSat revisited this source in 2016  May 6-7 $\sim$ 20 years after its RXTE observations. We present results from broadband (0.3-40 keV) study of the source using data from AstroSat observations with the SXT and the LAXPC detectors. Results on the spin period, energy dependent pulse profile, and pulsed fraction of the source are presented in this paper. We have also analyzed the broadband spectrum of the source in a 0.4-20 keV X-ray band and report the detection of a broad Iron line at 6.3 keV in its spectrum. The paper is structured as follows: introduction in section 1 is followed by a description of observation and data reduction in section 2. In section 3, we present results from the timing and spectral analysis. Details of the timing analysis e.g.power density spectrum of the pulsar, spin period, pulse profiles and pulse fraction are reported in subsection 3.1. In section 4 we present results of the spectral analysis. Significance of the results and their implication on the source model are examined in the discussion section 5.


\section{Observations and Data reduction}
\label{sect:Obs}
The binary 3A 0726-260 was observed with the SXT \& LAXPC instruments on-board AstroSat Satellite on 2016 May 6 from 19:59:47.2 hh:mm:ss UTC to 2016 May 7 22:54:29.2 hh:mm:ss UTC, comprising of 14 orbits, for a total duration of $\sim$ 91 ksec. A description of the AstroSat Satellite and its instrument is provided in \citet{2006AdSpR..38.2989A} and \citet{2014SPIE.9144E..1SS}.

The LAXPC instrument consists of 3 identical collimated detectors (LAXPC 10, LAXPC 20 and LAXPC 30)  covering 3-80 keV energy range, having 5 anode layer geometry with 15 cm deep X-ray detection volume providing an effective area of about 6000  cm$^{-2}$ at 10 keV. LAXPCs are filled with Xenon-Methane (90\% :10\%) mixture at 1520 torr and have a field of view of 0.9$^{\circ}$$\times$0.9$^{\circ}$. Every valid photon event detected in the LAXPC detectors is time tagged to an accuracy of 10$\mu$s. Detailed description of the characteristics of the LAXPC instrument can be found in \citet{2016SPIE.9905E..1DY}, \citet{2017JApA...38...30A}, \citet{2016ExA....42..249R} and calibration details in \citet{2017ApJS..231...10A}.

Level 1 data are converted to Level 2 data using LAXPC pipeline laxpcsoft\footnote{http://astrosat-ssc.iucaa.in/?q=laxpcData} version 2018 May 19. Level-2 data contain (i) light curve in broad band counting mode (modeBB) and (ii) event mode data (modeEA) with information about the arrival time, pulse height and the layer of origin of each detected X-ray (iii) housekeeping data and parameter files are stored in mkf file. The laxpc pipeline removes the overlapping data between consecutive orbits. Good time intervals were generated by filtering out Earth occultation and passages of South Atlantic Anomaly (SAA) regions. This pipeline software has individual routines to extract light curves and background lightcurves. We have used event mode data (ModeEA) for our analysis. This code generates a background model that is based on the observations of source-free sky regions. For our work, we have used data from only LAXPC 10, and LAXPC 20 units as the gas gain of LAXPC 30 was not stable due to gas leakage \citep{2017ApJS..231...10A}. The LAXPC source and background light curves are generated using the task 'laxpc\_make\_lightcurve' and 'laxpc\_make\_backlightcurve' tasks of LAXPCsoftware.
We generated energy dependent lightcurves in the energy bands 3-5 keV, 5-8 keV, 8-20 keV and overall 3-20 keV using data only from the top layer of LAXPC as most incident photons ($\sim$ 90\%) of $<$ 20 keV energy are absorbed in the top layer comprising Anode 1 and Anode 2. Further, the lightcurves in 20-40 keV and 3-40 keV are generated from both the LAXPCs 10 and 20, from the top layer only, as including other layers increased the background counts which dominated the source photons.
For the timing analysis, combined data from the 2 LAXPCs were used. For the spectral analysis, we have used data obtained only from LAXPC 10 \& LAXPC 20 separately with 512 and 256 channels binning as suggested by \citep{2017ApJS..231...10A} and used single routine LAXPC pipeline laxpcsoftv3.0.

SXT onboard AstroSat is a soft X-ray telescope sensitive in the 0.3-8 keV range. The effective area of SXT is $\sim$ 90 cm$^{2}$ at 1.5 keV. A detailed description of the SXT instrument can be found in \citet{2016SPIE.9905E..1ES, 2017JApA...38...29S}. The SXT observations of 3A 0726-260 were in the Photon counting mode. Full frame readout time resolution of the detector is $\sim$ 2.4 sec in SXT photon counting mode. The SXT level 1 data from 14 orbits were processed using SXTPIPELINE version AS1SXTLevel2-1.4b\footnote{http://www.tifr.res.in/~astrosat\_sxt/sxtpipeline.html} released on 2019 January 3, to generate level2 data for each orbit. Level 2 SXT data is filtered out from contamination by the charged particles from excursions through the SAA region and the Earth occultation. All the events with grade $>$ 12 were removed in the level 2 data. Merged event files from all the individual orbit were produced using python script sxt\_gti\_corr\_evt\_merger\_v05.py \footnote{http://www.tifr.res.in/~astrosat\_sxt/dataanalysis.html}. A circular region with 8 arcmin radius centered on the source and another circular region of 8 arcmin radius away from the source location, were used to extract the lightcurves and spectra for the source and background, respectively. The light curve and spectrum were generated using XSELECT under HEASoft (v 6.25).

A barycentric correction was applied on the lightcurves to incorporate the effect of Earth and satellite motion relative to the barycenter of the solar system. For the timing analysis of LAXPC, event files are barycenter corrected using AstroSat's barycentric correction tool as1bary\footnote{http://astrosat-ssc.iucaa.in/?q=data\_and\_analysis}. We have used ftool earth2sun\footnote{https://heasarc.gsfc.nasa.gov/ftools/fhelp/earth2sun.txt} for barycenter correction of the SXT lightcurve. The data were analyzed using HEASOFT 6.25\footnote{https://heasarc.gsfc.nasa.gov/lheasoft/download.html}. HEASOFT consists of (mainly) FTOOLS for general data extraction and analysis, XRONOS \citep{1992EMIS..59...59S} for the timing analysis, and XSPEC package \citep{1996ASPC..101...17A} for the spectral analysis. Table \ref{tab:count rates} lists details of count rates of 3A 0726-260 observed from the SXT and LAXPC detectors.

\begin{figure}[!ht]
\centering
\includegraphics[angle=-90, scale=0.5]{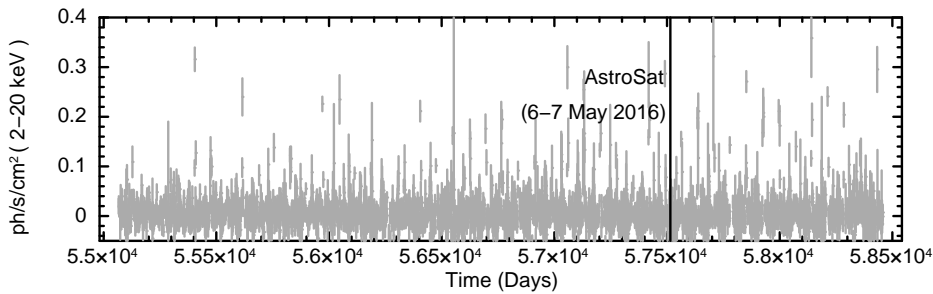}
    \caption{Maxi one day binned light curve in 2-20 keV energy band from MJD 55066.5 to MJD 58457.5 of the source 3A 0726-260 is shown in this plot. AstroSat observation on MJD 57514.83-57515.95 (2016 May 6-7) is indicated with a thick black vertical line.}
\label{fig1}
\end{figure}

\section{Data Analysis}
\label{sect:dataana}
\subsection{Timing Analysis}
One day binned MAXI light curve of the source from  MJD 55066.5 to MJD 58457.5, in 2-20 keV along with the AstroSat observation on MJD 57514.83 indicated with a thick black vertical line, is shown in Figure \ref{fig1}. As seen from the MAXI\footnote{http://maxi.riken.jp/top/slist.html} light curve, the AstroSat observation was made when the source was in a quiescent state.  
\begin{figure}[!ht]
\centering
\includegraphics[width=\columnwidth, height=12cm, scale=0.7]{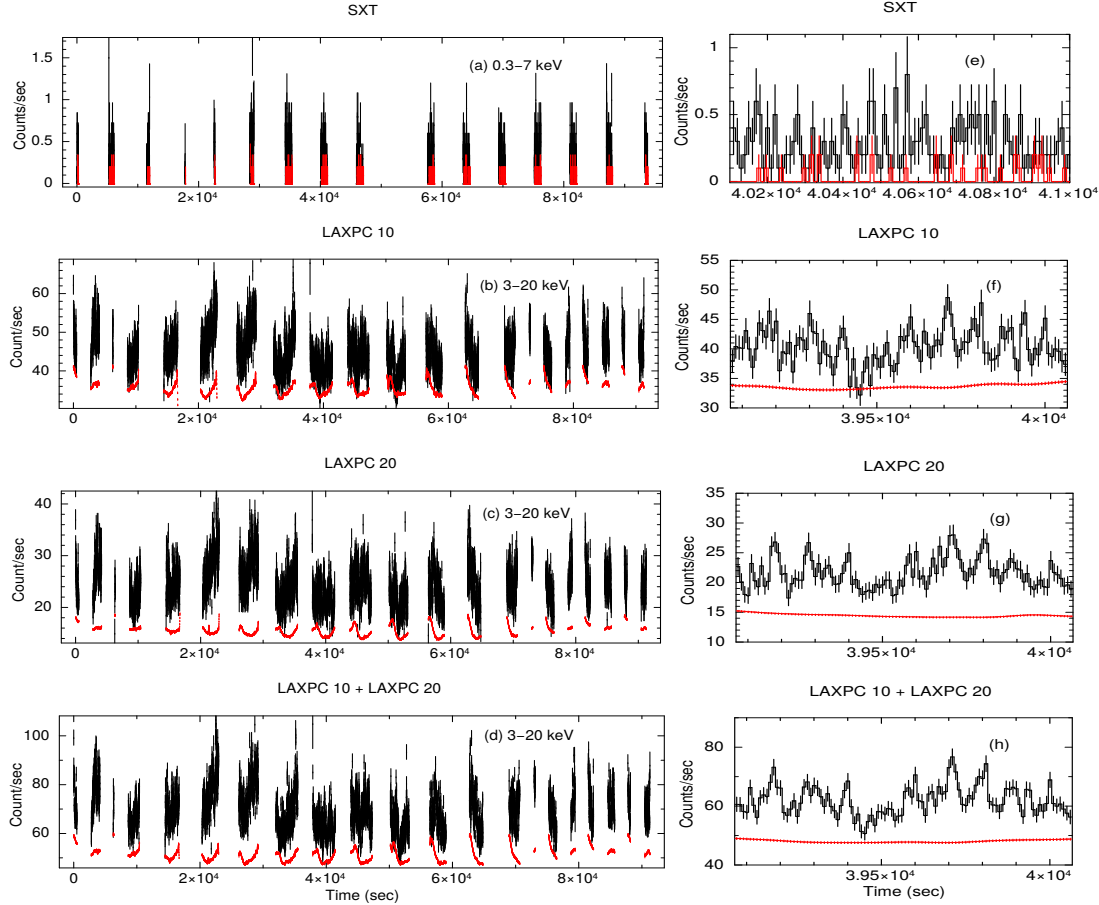}
    \caption{Top panel (a) shows the SXT lightcurve of 3A 0726-260 in 0.3-7 keV energy range. Panel (b) shows the 3-20 keV LAXPC 10 lightcurve, and panel (c) shows the LAXPC 20 lightcurve. The bottom panel (d) shows the combined LAXPC 10 and LAXPC 20 lightcurves. All the barycenter corrected lightcurves shown in this plot are in 10 sec bins. Estimated background rates for the detector are presented in the same panel with red color. The gaps in the light curve are due to the passage of the satellite through the South Atlantic Anomaly regions. Right side panels (e), (f), (g), and (h) shows the zoomed in 10 seconds binned lightcurve of the left side panel of SXT, LAXPC 10, LAXPC 20 and combined LAXPC 10 \& 20 lightcurves respectively. These zoomed in lightcurves show $\sim$ 103 sec periodicity of the source. The LAXPC error on the background count rate (presented in red color) is 2\% systematic errors.}
\label{fig2}
\end{figure}

Source and background light curves were extracted from the data of 14 orbits in 3-20 keV with a time resolution of 1 second for the LAXPCs, and a time resolution of 2.377501 seconds for the SXT in 0.3-7 keV. The X-ray light curves of 10 sec binning are presented in Figure \ref{fig2} for the SXT, LAXPC 10, LAXPC 20, and combined rates of the two LAXPCs. Estimated background rates are also shown in red color in the same panels. The systematic error of 2\% is added to the background lightcurve as suggested by the background light curve analysis task 'laxpc\_make\_backlightcurve' of LAXPCsoftware(Format A)\footnote{http://astrosat-ssc.iucaa.in/?q=laxpcData}.

\begin{figure}[!ht]
\centering
\includegraphics[width=\columnwidth, angle=-90, scale=0.4]{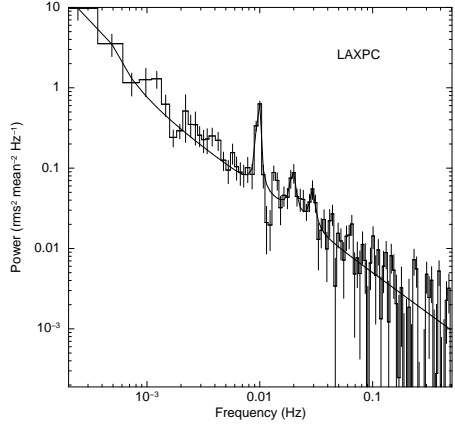}
\caption{The power density spectrum (PDS) of 3A 0726-260 generated using the combined LAXPC 10 and LAXPC 20 data from 1 sec binned barycenter corrected lightcurve in 3-20 keV, is shown in this figure. A strong pulsation peak at $\sim$ 9.7$^{+0.03}_{0.03}$ mHz is observed in the PDS. Harmonics of the pulse peak at 0.2$^{+0.01}_{-0.01}$ mHz \& 0.3$^{+0.01}_{-0.01}$ mHz are also observed in the same PDS.}
\label{fig3}
\end{figure}

Power density spectrum (PDS) of the source is generated using the 1 second binned, combined lightcurves of LAXPC 10 and LAXPC 20 using ftool "powspec" in HEASOFT. The light curve was divided into stretches of 4096 bins per interval. The PDS from all the segments were averaged to produce the final PDS for the observation. Poissonian noise was subtracted from the PDS, and they were normalized such that their integral gives the squared rms fractional variability normalized to units of (rms/mean)$^{2}$/Hz. PDS generated from 0.2 mHz to 0.5 Hz showed a very strong peak pertaining to the spin period of the source at 9.7$^{+0.03}_{0.03}$ mHz and two harmonics at 0.2$^{+0.01}_{-0.01}$ mHz \& 0.3$^{+0.01}_{-0.01}$ mHz as seen in Figure \ref{fig3}. 
An estimate of the pulsation period with better precision is achieved by using the standard $\chi$$^{2}$ maximization technique with the ftools task "efsearch".

\begin{figure}[!ht]
\centering
\includegraphics[width=\columnwidth,angle=0, scale=0.7]{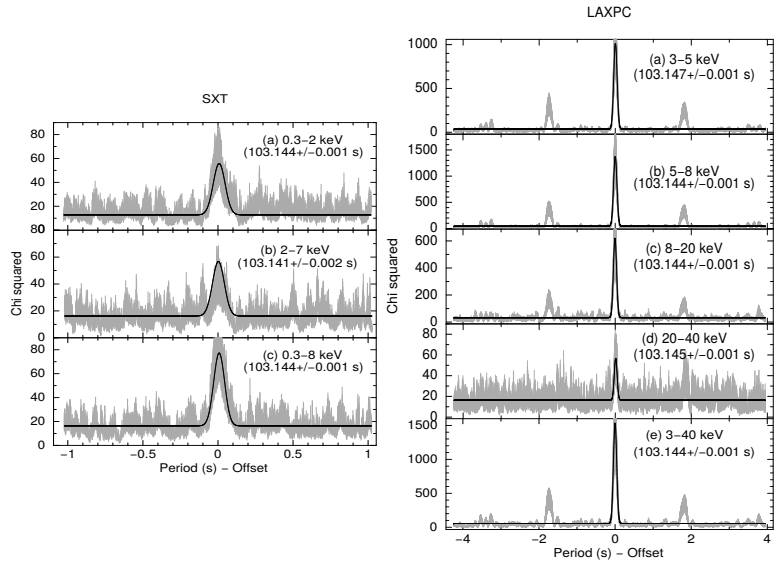}
\caption{The best fit pulse period of the pulsar 3A 0726-260 estimated using ftool "efsearch" from SXT and combined LAXPC 10 and LAXPC 20 light curves. The left side panel shows the best pulsation period estimated from the source in different energy intervals, using 2.377501 sec binned barycenter corrected SXT lightcurve. The right side panel shows the best pulsation period derived from barycenter corrected 1 sec binned LAXPC light curves.}
\label{fig4}
\end{figure}

\begin{table*}
	\centering
	\caption{SXT and LAXPC average count rates in different energy bands for 3A 0726-260 and background. The error on the LAXPC background count rate is 2\% systematic errors. Pulsation period and pulsed fraction in different energy intervals of SXT and LAXPC are presented in this table.}
	\label{tab:count rates}
	\begin{tabular}{lcccc} 
		\hline
		Observations &Energy & Average &Pulsation&Pulsed\\
			     &Range  & Count&Period&Fraction\\
			     &(keV)  &(counts/sec)& (sec)&(\%)\\
		\hline
\hline
\multicolumn{5}{c}{SXT}\\
\hline
		Source (+Background)	& 0.3-2   & 0.13$\pm$0.003	&103.144$\pm$0.001 & \\
					& 2-7	  & 0.11$\pm$0.003	&103.141$\pm$0.002 & \\
  		Background		& 0.3-2	  & 0.02$\pm$0.001	&-& \\
					& 2-7     & 0.01$\pm$0.001	&-& \\
		Source (-Background)	& 0.3-2	  & 0.11$\pm$0.004	&-&42.2$\pm$10.0 \\
					& 2-7	  & 0.10$\pm$0.003	&-&26.8$\pm$10.0 \\
\hline
\multicolumn{5}{c}{LAXPC}\\
\hline
		Source (+Background)	& 3-5    & 16.6$\pm$0.02	&103.147$\pm$0.001 & \\
					& 5-8	 & 18.0$\pm$0.02	&103.144$\pm$0.001 & \\
					& 8-20	 & 35.0$\pm$0.03	&103.144$\pm$0.001 & \\
					& 3-20   & 69.3$\pm$0.04	&103.144$\pm$0.001 & \\
					& 20-40  & 33.0$\pm$0.03        &103.145$\pm$0.001 & \\ 
	
  		Background		& 3-5 	 & 11.0$\pm$0.2		&-& \\
					& 5-8    & 11.3$\pm$0.2		&-& \\
					& 8-20   & 26.0$\pm$0.5		&-& \\
 		 			& 3-20 	 & 51.2$\pm$1.0		&-& \\
					& 20-40  & 31.2$\pm$0.6         &-&\\

		Source (-Background)	& 3-5	 & 5.6$\pm$0.2		&-&16.0$\pm$1.1 \\
					& 5-8 	 & 6.7$\pm$0.2		&-&14.8$\pm$1.0 \\
					& 8-20	 & 8.7$\pm$0.5		&-&14.2$\pm$1.0 \\
					& 3-20	 & 18.1$\pm$1.0		&-&13.4$\pm$0.7 \\
					& 20-40  & 1.5$\pm$0.6		&-&23.0$\pm$6.0 \\

\hline
\end{tabular}
\end{table*}

\begin{figure*}[!ht]
\centering
\includegraphics[width=12 cm, scale=0.9]{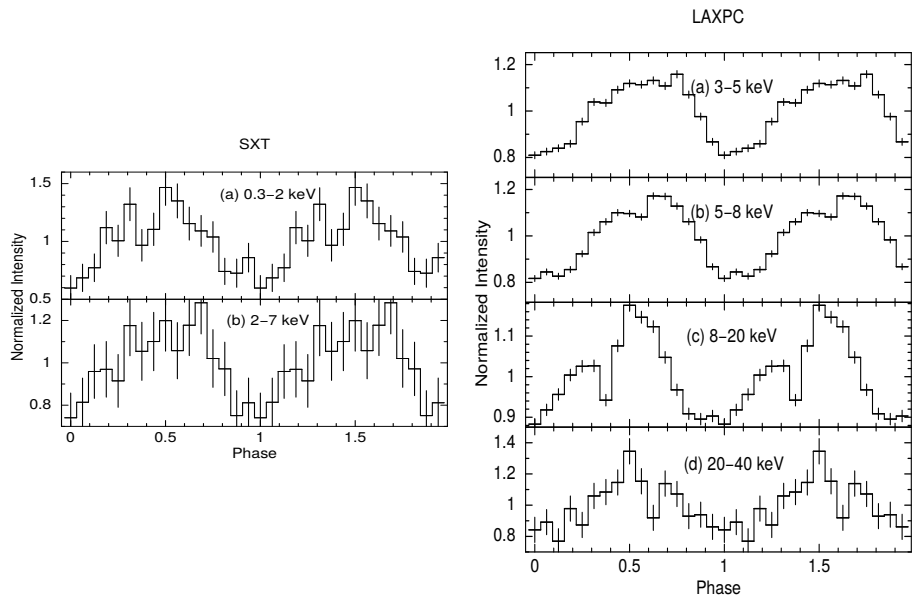}
\caption{The energy evolution of pulse profile generated using ftool "efold" is shown in this figure. Left panels shows pulse profile in (a) 0.3-2 keV, and (b) 2-7 keV generated from 2.377501 sec binned SXT light curves. And Right panel show pulse profiles generated from LAXPC light curves in, (a) 3-5 keV, (b) 5-8 keV, (c) 8-20 keV, and (d) 20-40 keV. These single and double peaked pulse profiles have been generated by folding with their best fit pulse periods and 16 phase bins/period.}
\label{fig5}
\end{figure*}

The pulsation period is searched around the approximate period 102.89 sec, as observed in the PDS generated from the combined LAXPC light curve. The 1 sec binned light curve is folded with 8192 different periods around the approximate period with 1 ms resolution and 16 phase bin per period. The distribution between the maximum $\chi$$^{2}$ and the pulsar period shows a Gaussian profile indicating the detected pulsation period. This pulsation peak was fit with a Gaussian function to estimate the pulse period. The error on the Gaussian center is the error in the observed pulsation period. 
Using the same methodology for the estimation of the pulsation period, as done with the LAXPC lightcurve, the pulse period from the SXT data is estimated from the 2.377501 sec binned light curve of SXT using 2048 different periods around the approximate period. We were unable to use 8192 different period searches around approximate period as SXTs bin time is 2.3775 sec, unlike 1 sec binning of LAXPC lightcurves. The pulsation period value is shown in different energy bands for SXT and LAXPC in left panels and right panels of Figure \ref{fig4}. The best estimated pulse period from all the observations are tabulated in Table \ref{tab:count rates}.

Pulse profiles in different energy bands were generated by folding the lightcurve using the ftool "efold" over the exact pulsation period 103.144 sec obtained by standard $\chi$$^{2}$ maximization technique of LAXPC lightcurve. The pulse profiles have been generated with 16 phase bins per period. The pulse profiles in 0.3-2 keV and 2-7 keV for background subtracted SXT lightcurves are shown in the left panels of Figure \ref{fig5}. The pulse profiles in 3-5 keV, 5-8 keV, 8-20 keV, and 20-40 keV background subtracted lightcurves of LAXPC are shown in the right panels of Figure \ref{fig5}. The pulse profile in 0.3-2 keV, 2-7 keV from the SXT, and 3-5 keV and 5-8 keV from the LAXPCs show similar shape with a broad pulse. The pulse shape in 8-20 keV and 20-40 keV appears different with a double hump pulse profile. This suggests that the pulse shape is energy dependent. The pulse fraction of pulse profiles is calculated from the relation (F$_{max}$-F$_{min}$)/(F$_{max}$+F$_{min}$), where F$_{max}$ and F$_{min}$ are the maximum and minimum values of the observed photon flux \citep{2011A&A...526A...7N}. It will be noticed that the pulse fraction in 3-20 keV inferred from the LAXPCs is constant at about 15\%. The 2-7 keV pulse fraction from SXT is 26.0$\pm$10, and considering the large error, it is also consistent with a mean pulse fraction of 15\%. The pulse fraction in 0.3-2 keV seems to suggest a larger value, but given the large uncertainty associated with it, no definite conclusion can be drawn about its energy dependence.

\subsection{Spectral Analysis}
We have performed a combined broadband spectral analysis of SXT in 0.4-7 keV and LAXPC units 10 and 20 in 3-20 keV. LAXPC spectrum is extracted from the top layer (Anode 1 and 2) data of the merged 14 LAXPC orbits. Spectrum is extracted from the top layer to optimize the inclusion of X-ray photons and reduce the background. The LAXPC software having a single routine to the extract spectra, background spectra, and response matrix files, was used\footnote{http://astrosat-ssc.iucaa.in/?q=laxpcData}. The script 'laxpcl1.f' provided with the software processes the level-1 orbit data to generate the level-2 standard products like the event file, spectrum for source, and background. To correct for the drift in the gain of the detectors with time, we have used the script 'backshift.f'provided with the software. This code generates background model which is based on the observations of source-free sky regions. The background LAXPC data closest to gain is chosen.

\begin{figure*}[!ht]
\centering
\includegraphics[width=\columnwidth, angle=0, scale=0.6]{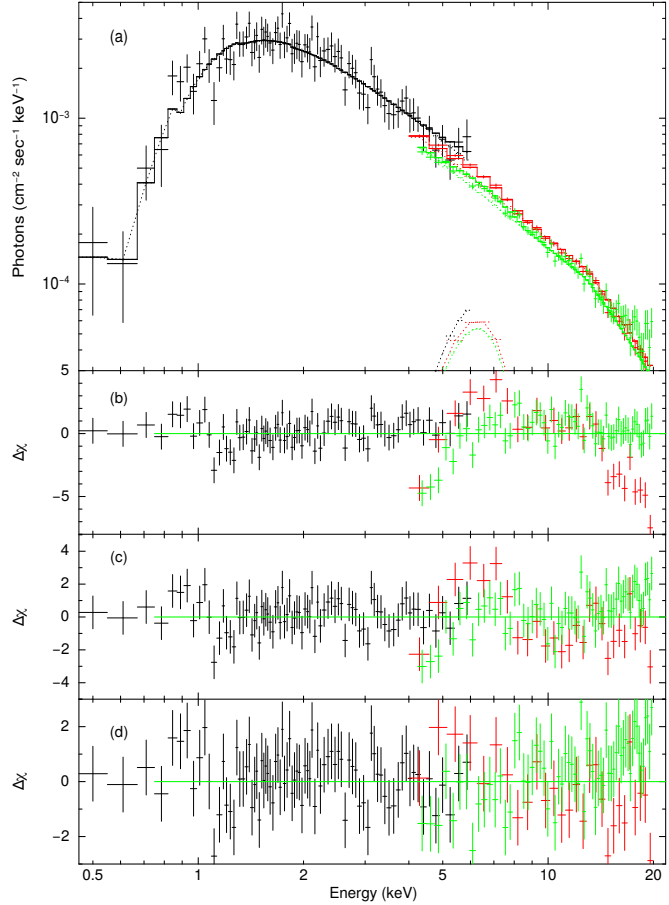}
\caption{Panel (a) above shows the unfolded combined energy spectrum. The  SXT (0.4-7 keV) derived spectrum is shown in black, and the spectra from  LAXPC 10 and 20 (4-20 keV) are shown in green and red color, respectively. The best fit models is shown with a solid line. Panel(b) shows the residual for the absorbed powerlaw (TBabs*powerlaw) model. Panel (c) shows the residual after adding the highecut model inbuilt XSPEC. Panel (d) shows the residual after adding the Gaussian line due to Fe K$\alpha$ line at 6.3$^{+0.3}_{-0.4}$ keV. }
\label{fig6}
\end{figure*}

A gain correction is applied to the SXT spectra by freezing gain slope to value 1\footnote{http://astrosat-ssc.iucaa.in/uploads/sxt/readme\_sxt\_arf\_data\_analysis.txt, https://www.tifr.res.in/~astrosat\_sxt/instrument.html} and the gain offset was allowed to vary with the fit as suggested by the SXT instrument team. We obtained an offset of -0.12$\pm$0.02. The SXT spectrum of the Pulsar was fitted with an absorbed powerlaw model that included an interstellar absorption TBabs \citep{2000ApJ...542..914W} and powerlaw \footnote{https://heasarc.gsfc.nasa.gov/xanadu/xspec/manual/node211.html} listed in XSPEC. The value of column density is fixed at the N$_{H}$ value obtained from SXT 0.4-7 keV energy spectrum, N$_{H}$=0.75$\times$ 10$^{22}$cm$^{-2}$ for the combined SXT and LAXPC spectral fitting. The combined spectrum gave a poor fit with $\chi^{2}$/dof value of 478/171 with 2\% systematics used for spectral fitting.

\begin{equation}
f(E)=KE^{-\Gamma}\times \left\{ \begin{array}{ll}
1 & \mbox{($E \leq E_{c}$)}\\
exp^{-(E-E_{c})/E_{f}} & \mbox{($E > E_{c}$),} \end{array} \right.
\label{equ1}
\end{equation}
where, K is normalization factor (photons/keV$^{-1}$cm$^{-2}$s$^{-1}$ at 1keV), $\Gamma$ is photon index of powerlaw, E is the photon energy, $E_{c}$ is the cut off energy in keV and $E_{f}$ is the e-folding energy in keV.

\begin{table*}[!ht]
	\centering
	\caption{Details of best fit spectral parameters of combined SXT (0.4-7 keV), LAXPC 10 and LAXPC 20 (4.0-20 keV) energy spectra. }
	\label{tab2}
	\begin{tabular}{lll} 
		\hline
		Model & Parameters & Values \\

		\hline

\multicolumn{3}{c}{Combined SXT and LAXPC}\\
\hline
\hline
  		TBabs			& nH (10$^{22}$  cm$^{-2}$)			& 0.75 (Fixed)					\\
		Powerlaw		& PhoIndex ($\Gamma$)				& 1.7$^{+0.03}_{-0.03}$ 			\\
 		Highecut		& cutoffE (keV)					& 12.9$^{+0.7}_{-0.7}$			\\
					& foldE (keV)					& 10.9$^{+2.4}_{-2.0}$			\\
		Gaussian		& LineE (keV)		       			& 6.3$^{+0.3}_{-0.4}$ 			\\
					& Sigma (keV)	              			& 1.1$^{+0.4}_{-0.4}$ 			\\
					& Norm ($\times$10$^{-4}$)			& 1.9$^{+0.8}_{-0.7}$ 			\\
		Constant		& SXT						& 1.0 (Fixed)		 			\\
					& LAXPC 10					& 0.74$\pm$0.04  				\\
					& LAXPC 20					& 0.84$\pm$0.04  				\\
		Reduced $\chi^{2}$ (dof)& Without highecut \& gaussian			& 2.8 (171)					\\
 					& Without gaussian				& 1.6 (169)					\\
		Reduced $\chi^{2}$ (dof)& Best fit			                & 1.3 (166)					\\
		Flux (0.4-20 keV)	& ergs cm$^{-2}$ sec$^{-1}$ ($\times$10$^{-11}$)& 8.6 $^{+0.3}_{-0.3}$				\\
		Flux (0.4-7 keV)	& ergs cm$^{-2}$ sec$^{-1}$ ($\times$10$^{-11}$)& 6.6 $^{+0.3}_{-0.2}$ 				\\
		Flux (4-20 keV)		& ergs cm$^{-2}$ sec$^{-1}$ ($\times$10$^{-11}$)& 3.9 $^{+0.3}_{-0.2}$ 				\\
\hline
\end{tabular}
\end{table*}

Further, a high energy cutoff \citep{1983ApJ...270..711W} model available in XSPEC is added to the absorbed powerlaw (see equation \ref{equ1}). We obtained a significant improvement in the fit parameters with $\chi^{2}$/dof=1.59 (268/169).
This model leaves a residual at $\sim$ 6.4 keV. Inclusion of a Gaussian line at 6.3 keV due to iron K$\alpha$ line, significantly improved the fit to $\chi^{2}$/dof=1.34 (223/166). Total flux and its error have been obtained using "cflux" model in XSPEC. Total flux observed in 0.4-20 keV is inferred as 8.6 $^{+0.3}_{-0.3}$  $\times$ 10$^{-11}$ ergcm$^{-2}$s$^{-1}$. The best fit spectral parameters are tabulated in Table \ref{tab2}. Figure \ref{fig6} shows the combined SXT and LAXPC unfolded spectrum, and the solid line represents the combined best fit model.

\section{Results and Discussion}
\label{sect:ResDis}

The spectral and timing characteristics of the Be/X-ray binary Pulsar 3A 0726-260 have been determined from AstroSat, SXT and LAXPC data of 2016 May 6-7 (MJD 57514.83). The observations below 2 keV from SXT are reported for the first time. \citet{1984ApJ...280..688S}, obtained the first X-ray light curve using the SSI instrument on the Ariel V satellite. \citet{1997ApJ...489L..83C} discovered that the X-ray source in 3A 0726-260 was a pulsar with a spin period of 103.2$\pm$0.2 sec. The AstroSat observations 20 years later give a spin period of 103.144$\pm$0.001 sec, indicating that if at all, there has been only a marginal change in the spin period. \citet{1997ApJ...489L..83C} did not find any clear periodicity in 20-40 keV PCA lightcurve. However, due to the higher effective area of LAXPC detectors in the 20-40 keV  band \citep{2006AdSpR..38.2989A}, we could detect weak pulsations in the LAXPC data for the first time from the source in 20-40 keV range. Due to large errors in the 1996 RXTE measurement, it is difficult to derive the rate of variation in the spin period. A crude estimate of the spin-up rate can be made from the two measurements spaced 20 years apart, which gives 8.9 $\times$10$^{-11}$ s/s, and it can be treated as an upper limit. It may be noted from the MAXI light curve that despite being a Be binary, there is no indication of any major outburst in 3A 0726-260 over almost 10 years. This suggests that this pulsar is accreting at almost a steady accretion rate with no indication of any instability leading to spike in the accretion rate and occurrence of any outburst. The X-ray pulse profile shows a single peak structure upto 5 keV. In the pulse profiles above 5 keV, there is a hint of change from a single peaked profile to a double peak structure. Above 8 keV appearance of the double-peak in the profile is unmistakable, as seen in Figure \ref{fig5}. We observed from Figure \ref{fig5} that in the double peak pulse profile above eight keV, the first peak is less prominent compared to the second peak whereas, above 20 keV, the first peak becomes stronger compared to the second peak. The estimated values of the pulse fraction in different energy bands are presented in Table \ref{tab:count rates}. The change in the pulse profile from a single peak to a weak double peaked structure may be explained by intrinsic change occurring in the beaming pattern from a pencil beam to a fan beam which results in the beam to move out of our line of sight as observed from the X-ray binary Pulsar LMC X-4 \citep{2010ApJ...720.1202H}. The change in the pulse profile can also be attributed to a transition in accretion pattern from a smooth accretion stream at low energies to several narrow accretion streams at high energy that are phase-locked with the neutron star as observed from EXO 2030+375 \citep{2015RAA....15..537N}. In this work, we deduce the X-ray spectrum in 0.4-20 keV using combined SXT and LAXPC data. The source flux during the Astrosat observation in 2-20 keV is estimated as 5.7 $^{+0.3}_{-0.3}$$\times$10$^{-11}$ ergs cm$^{-2}$ sec$^{-1}$. This is in agreement with the flux value 6.9$\times$10$^{-11}$ ergs cm$^{-2}$ sec$^{-1}$ reported by \citet{1997ApJ...489L..83C}, from RXTE/PCA observation of the source on 1997 June 7 (MJD 50606). The power-law spectrum is steeper with photon index $\Gamma$=1.7$^{+0.03}_{-0.03}$ compared to the RXTE/PCA value. In the spectrum, there is no indication of the presence of any cyclotron line feature in the spectrum. \citet{1997ApJ...489L..83C} reported a hint of an iron line at $\sim$ 6.6 keV and width $\sim$ 150 eV. In this work, we clearly detect the presence of an of broad 1.06$^{+0.4}_{-0.4}$ keV, iron K$\alpha$ line at $\sim$ 6.3 keV in the combined SXT and LAXPC spectrum. The broad iron line feature may be due to the presence of two lines at $\sim$ 6.4 keV and $\sim$ 6.6 keV, which could not be clearly resolved by LAXPC detectors. Hopefully, more spectral studies in the future will resolve this ambiguity.

\begin{acknowledgements}
We are thankful to the reviewer for the constructive comments, which improved the manuscript. This publication uses the data from the AstroSat mission of the Indian Space Research Organisation (ISRO), archived at the Indian Space Science Data Centre (ISSDC). This work has used the AstroSat data from the Large Area X-ray Proportional Counter (LAXPC) detectors developed at TIFR, Mumbai. We would like to thank LAXPC team  for verifying and releasing the data via the ISSDC data archive and providing the necessary software tools. This work has used the data from the Soft X-ray Telescope (SXT) developed at TIFR, Mumbai, and the SXT POC at TIFR is thanked for verifying and releasing the data via the ISSDC data archive and providing the necessary software tools. This paper makes use of our proposed data from the AstroSat mission of the Indian Space Research Organisation (ISRO), archived at the Indian Space Science Data Centre (ISSDC). JR would like to thank ISRO for providing funding support. JR would like to thank IUCAA for providing facilities. Author B. Singari contributed in the work during his two months visit at UM-DAE CEBS as a SARP student. JR thanks Prof. A. R. Rao for valuable suggestions and discussions. This research has made use of softwares obtained through the HEASARC Online Service, provided by the NASA/GSFC, in support of NASA High Energy Astrophysics Programs. This research has made use of MAXI data provided by RIKEN, JAXA and the MAXI team.  

\end{acknowledgements}

\bibliographystyle{raa}
\bibliography{main.bib}

\end{document}